\documentclass[11pt,letterpaper]{article}
\usepackage[letterpaper,margin=1in]{geometry}
\usepackage{mathptmx}
\usepackage[T1]{fontenc}
\usepackage[utf8]{inputenc}
\usepackage{graphicx}
\usepackage{booktabs}
\usepackage{tabularx}
\usepackage{array}
\usepackage{caption}
\usepackage{float}
\usepackage{microtype}
\usepackage{titlesec}
\usepackage{fancyhdr}
\usepackage[dvipsnames]{xcolor}
\usepackage[colorlinks=true,allcolors=NavyBlue]{hyperref}
\usepackage{url}
\usepackage[authoryear,round]{natbib}
\newcolumntype{Y}{>{\raggedright\arraybackslash}X}
\titleformat{\section}{\large\bfseries}{\thesection}{0.6em}{}
\titleformat{\subsection}{\normalsize\bfseries}{\thesubsection}{0.6em}{}
\titlespacing*{\section}{0pt}{1.6ex plus .5ex}{1ex}
\titlespacing*{\subsection}{0pt}{1.3ex plus .4ex}{0.8ex}
\title{\vspace{-0.5em}\textbf{Scaling Peer Assessments: An Integrity Report from a Large Engineering Internship}}
\author{%
  \textbf{Jinal Gupta}$^{1,*}$ \quad Pavani Ayinampudi$^{1}$ \quad Aditya B.M.V.$^{1}$ \quad Prakash Hegade$^{1}$\\
  Rohit Sharma$^{2}$ \quad Sakshi Sharma$^{2}$ \quad Meenakshi V$^{2}$ \quad S.R.S. Iyengar$^{2}$\\[0.8em]
  {\small\itshape $^{1}$ANNAM.AI, Indian Institute of Technology Ropar, Rupnagar, Punjab, India}\\
  {\small\itshape $^{2}$Indian Institute of Technology Ropar, Rupnagar, Punjab, India}\\[0.5em]
  {\small $^{1}$\{pavania.harvard2025, adityabmv, prakash.hegade\}@gmail.com}\\
  {\small $^{2}$\{sakshi.23csz0006, rohit.24csz0014, meenakshi.19csz0013, sudarshan\}@iitrpr.ac.in}\\
  {\small $^{*}$Corresponding author: jinalbirla@gmail.com}
}
\date{}

\begin{document}
\maketitle
\thispagestyle{empty}

\begin{abstract}
Assessing learning in large classrooms presents a significant challenge for individual instructors, who may have limited capacity to evaluate the understanding, participation, and assessment behaviour of every student. Peer assessments have been a way of distributing this responsibility among learners, allowing them to evaluate and provide feedback to one another while reducing dependence on instructor-led assessments. Building on this approach, we implemented a peer validation model within a large, multi-institutional internship programme in which students who demonstrated sufficient understanding were authorised to assess and validate their peers through short oral discussions. The assessment process began with the instructor validating a small group of students, who were then authorised to validate their peers, allowing the process to gradually expand across the cohort and operate at scale. This study examines how participants experienced the model and the extent to which assessment integrity was maintained, using an end-of-programme survey of 238 consenting respondents. Most participants regarded the activity as worthwhile, with 79.8\% reporting that they solved problems they could not previously solve. However, 29.0\% acknowledged at least one instance of reduced effort, a lowered validation standard, or reciprocal validation, while 88.7\% believed that at least a little validation had occurred without proper examination. When asked how the process could be strengthened, participants selected post-validation discussion of solutions approximately twice as often as closer auditing or mentor-led validation. These findings provide descriptive evidence of both the potential and the integrity challenges of using peer validation as a scalable assessment approach in large learning environments.
\end{abstract}


\vspace{1em}

\section{Introduction}

Peer validation has long been explored as an approach to addressing the challenges of assessment in large classrooms. As classrooms grow in size, relying solely on instructors to evaluate the understanding of every learner becomes increasingly difficult. An instructor managing hundreds or thousands of students has limited time to individually examine conceptual understanding, provide meaningful feedback, and determine whether students have truly mastered the concept. Large-scale assessment therefore tends to rely on measurable indicators such as assignment completion, quiz scores, examination performance, or course completion. While these measures are useful for monitoring progress, they may not fully reflect the depth of a student's understanding.

Instructor-led assessment can also influence students' learning behaviour. When the instructor is the primary authority responsible for validating learning, students may become more focused on meeting assessment requirements, obtaining marks, or clearing a particular threshold than on developing a deeper understanding of the subject. The assessment process can consequently become associated with performance and evaluation rather than exploration, discussion, and learning. This is particularly challenging in large classrooms, where instructors have limited opportunities to interact individually with learners and to understand their specific gaps.

Peer validation offers a potential alternative by distributing part of the assessment responsibility among learners. Students can evaluate their peers, explain concepts, question reasoning, identify gaps in understanding, and provide immediate feedback. The process can benefit both participants, because assessing another learner requires the assessor to engage critically with the subject matter, while the learner receives an opportunity to articulate and defend their understanding. For instructors, peer assessment can also reduce the burden of individually evaluating large numbers of students and can foster a more participatory learning environment.

Extending peer assessment to a large scale, however, introduces important questions about reliability and academic integrity. Peer relationships may influence the assessment process, particularly when students know one another. Friendship, familiarity, reciprocity, or differences in individual judgement may produce inconsistent assessment standards, and a student may be less critical when evaluating a peer than an instructor would be. The central challenge is therefore not simply whether students can assess one another, but whether they can maintain a meaningful and consistent standard of validation when responsibility for assessment is distributed across a large learner population.

To examine this question in practice, we implemented a peer validation model within a large, multi-institutional internship programme involving students in the thousands. Rather than relying entirely on a limited number of instructors or mentors to validate every student's understanding, students who demonstrated sufficient understanding were given responsibility to assess and validate their peers. This created a distributed validation process in which students could progress from being assessed to becoming assessors themselves.

The present study examines how this model functioned in practice and how students perceived the peer validation process. Specifically, we investigate whether students followed the intended assessment standards, whether they believed their peers maintained similar standards, and what challenges and improvements they identified. By examining learner experiences within a large-scale implementation, this study seeks to provide empirical evidence on the feasibility, reliability, and challenges of peer validation as a scalable approach to validating learning.

\section{Related Work}

Peer assessment is among the more extensively studied practices in higher education. In an early synthesis, \citet{Topping1998} concluded that it can achieve adequate reliability and validity across a wide range of applications and can approximate the formative effects of instructor assessment. Subsequent meta-analytic work has refined this view. \citet{Falchikov2000} found that peer marks correspond closely to teacher marks, most strongly when judgements are holistic and criteria are well understood. \citet{Double2020}, synthesising fifty-four control-group studies, reported a small-to-medium positive effect on academic performance. A more recent meta-analysis reaches a broadly similar conclusion \citep{Li2020}. Importantly, the benefits are not confined to the student being assessed: the act of evaluating a peer is itself a learning activity for the assessor \citep{Lundstrom2009,Li2010,VanPopta2017}. This reciprocal benefit is central to the appeal of peer learning, in which a cohort functions as a shared instructional resource rather than a queue for scarce expert attention \citep{Boud1999,Boud2001}.

The mechanism underlying these benefits is reasonably well understood. Formative assessment is thought to develop the learner's own capacity to judge quality, which is what enables feedback to be acted upon rather than merely received \citep{Sadler1989,Nicol2006}. This capacity has more recently been described as feedback literacy: the dispositions and skills a learner needs to interpret and use feedback effectively \citep{Carless2018}. On this account, much of the value of peer assessment lies not in the mark exchanged but in the judging itself, since a student who evaluates a peer against a standard is practising the very capability the activity is intended to build.

The reassuring evidence, however, derives from a particular configuration. In most studies an instructor is present and moderates the process without performing every judgement; marks are numeric, permitting aggregation across raters and calibration against a reference; and cohorts are small enough for a sample of judgements to be inspected. These safeguards are part of the finding rather than incidental to it. \citet{Magin2001}, for example, found that reciprocity, the concern that students exchange favourable ratings, accounted for only about one per cent of rating variance, but this was observed in a setting with instructor oversight and multiple raters per submission.

As peer assessment is scaled, the configuration begins to matter. In the canonical large-scale study, \citet{Kulkarni2013} reported an encouraging correlation between peer and staff grades but also documented systematic bias: students rated their own work several percentage points higher than staff did and favoured peers from their own region, and providing graders with feedback on their bias reduced it. Two features of that work are relevant here. Staff grades remained available as a reference against which peer grades could be checked, and the judgements were numeric, so calibration was possible at all. The social configuration also matters: reviewing anonymity effects, \citet{Panadero2019} found that anonymous arrangements were associated with more critical feedback and more favourable perceptions. Within computing education, where peer review is common, \citet{Indriasari2020} found benefits widely reported but concluded that claims regarding learning outcomes require firmer empirical support.

Two observations follow that motivate the present study. First, almost all of this evidence concerns a numeric mark or written feedback and assumes a flat structure with an instructor positioned outside the process. A binary validation, granted in conversation by a peer who was recently validated, with the authority to validate propagating through the cohort from an expert origin, has little empirical description in engineering education, and the safeguards that keep peer grading dependable, namely aggregation, calibration, and moderation, all presuppose a numeric judgement that a validation does not produce. Second, the prevailing outcome measure in this literature is agreement between peer and instructor marks; where no instructor mark exists, that measure is undefined, and participant reports become one of the few available windows onto how the process is functioning. The remedies proposed in the literature, such as rubrics, calibration, anonymity, and audits, are moreover authored by instructors, and whether participants who have experienced the process would themselves choose such remedies has received little attention. This study takes up an implementation of the first kind, uses measures of the second kind, and puts the third question to participants directly.

\section{Methodology}

\subsection{Research questions}

The study addresses one principal research question, examined through four sub-questions.

\textbf{RQ.} What did participants report about the integrity of a cascading peer validation system operated at cohort scale, and what changes did they identify as improvements?

\textbf{RQ1.} How common were reported reductions in effort, a lowered validation standard, and reciprocal validation?

\textbf{RQ2.} How did participants' accounts of their peers compare with their accounts of their own conduct?

\textbf{RQ3.} Which changes did participants select, and did selection depend on whether they had reported a problem?

\textbf{RQ4.} Was difficulty in finding a validator associated with any of the reported problems?

\subsection{Context and the validation cascade}

The setting was a national, multi-institutional virtual internship programme. Its first phase admitted 1,036 engineering interns; the 583 who completed the foundational coursework advanced to a second phase, in which every intern undertook the validation activity reported here. Alongside project work, these interns worked through a set of structured problems that carried no marks. Progress was recorded through validation: an intern who could explain a problem and its underlying concepts to a validated peer, in a short oral discussion, was validated on that problem. Reciprocal validation, in which two interns validate each other, was prohibited by programme rule.

\begin{figure}[tbp]
\centering
\includegraphics[width=0.8\linewidth]{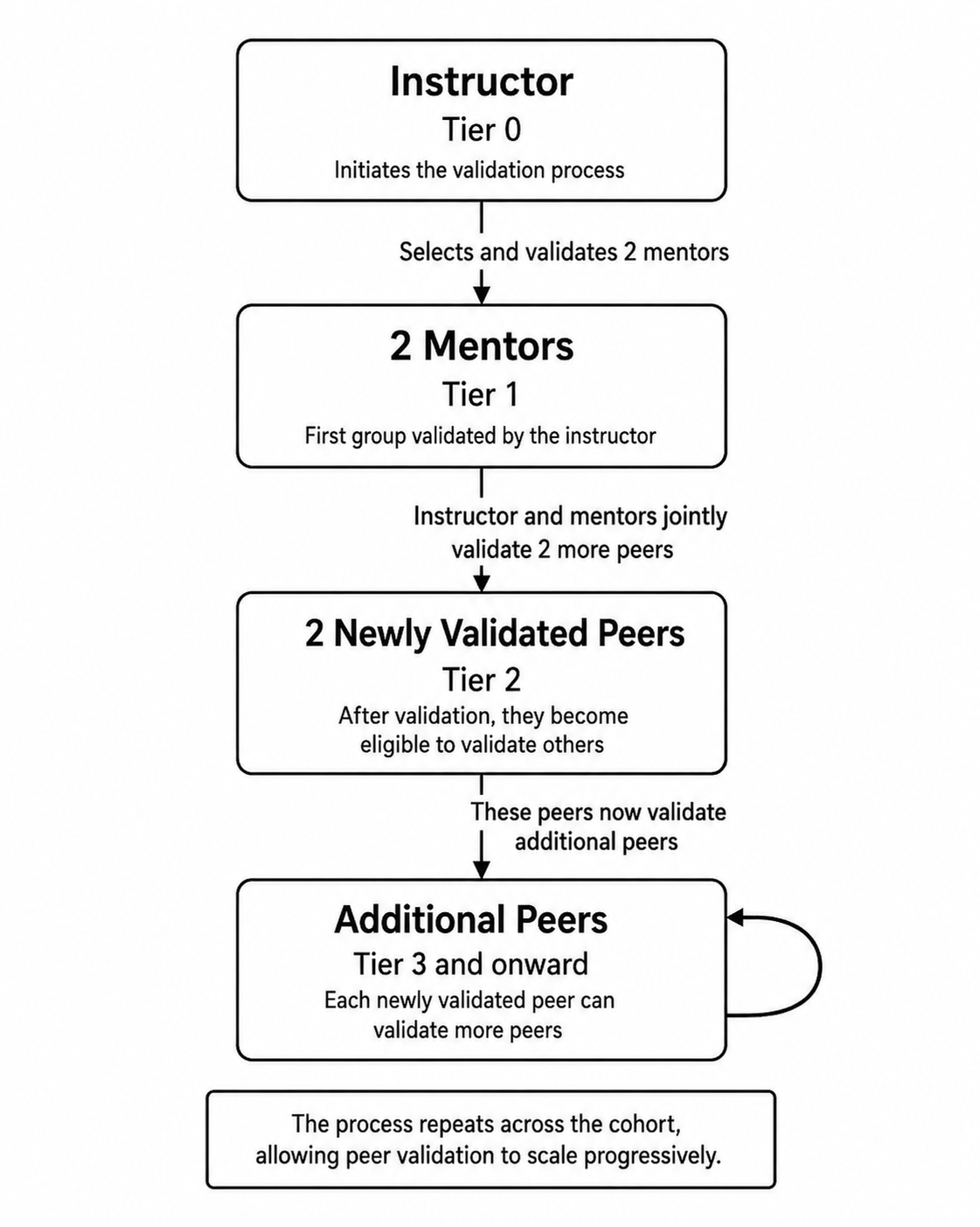}
\caption{The validation cascade. The instructor validated two mentors; those three validated two further validators. Thereafter, passing a validation discussion conferred the authority to validate other interns.}
\label{fig:cascade}
\end{figure}

Validating authority was distributed through the classroom rather than held centrally, as illustrated in Fig.~\ref{fig:cascade}. The instructor validated two interns as mentors; those three then validated two further interns, and from that point the process was self-propagating so that any intern who passed a validation discussion acquired the authority to validate others. There was no separate confirmation step for a routine validation. The platform awarded points for validation granted, recording how many interns each validator had validated. Some validations were subsequently invalidated and had to be repeated.

\subsection{Instrument and participants}

Data were obtained from an end-of-programme exit survey comprising forty items across eight sections. The section relevant to this study contained twelve items concerning peer validation: six five-point Likert items, two ordinal items, one multi-select item on possible changes, and free-text fields. Three Likert-scaled items were worded negatively, so agreeing with them means that the participant reported a problem.

Of the 314 surveys submitted, 241 respondents consented to research use, and only those records were analysed; 238 of them were routed to the validation section, the remaining three being on a track for which the section was not shown, as summarised in Table~\ref{tab:sample}. All figures reported below are based on 238 respondents, except for one item based on 237. Proportions are reported with Wilson 95\% confidence intervals. The multi-select item was analysed using Cochran's Q to test for an overall difference among options, followed by exact McNemar tests for paired comparisons; ordinal associations were assessed using Spearman's rho with a Holm correction across the family of ten tests; and subgroup contrasts used Fisher's exact test and are reported as exploratory. Free-text responses were anonymised before analysis. Analysis was conducted in Python using pandas, SciPy, and statsmodels.

\begin{table}[tbp]
\centering
\caption{Derivation of the analytic sample}
\label{tab:sample}
\begin{tabularx}{0.9\linewidth}{Y r}
\toprule
\textbf{Stage} & \textbf{n} \\
\midrule
Phase 1 intake & 1,036 \\
Advanced to Phase 2 (validation activity) & 583 \\
Exit surveys submitted & 314 \\
Consented to research use & 241 \\
\quad Consent withheld (excluded) & 73 \\
Routed to the validation section & 238 \\
\quad Not routed (section not shown) & 3 \\
\bottomrule
\end{tabularx}
\end{table}

\section{Results and Discussion}

\subsection{Results}

\begin{figure}[tbp]
\centering
\includegraphics[width=0.92\linewidth]{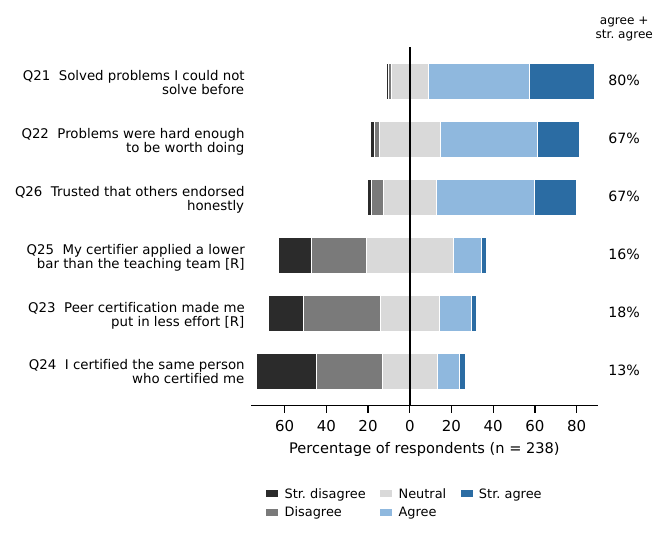}
\caption{Responses to the six peer validation Likert items (n = 238), centred on the neutral midpoint. Items marked [R] are reverse-worded and are shown in their original direction, so agreement indicates a reported problem.}
\label{fig:likert}
\end{figure}

The activity was regarded as worthwhile. Participants did not treat the exercise as busywork. Asked whether they had solved problems they could not previously have solved, 79.8\% agreed (95\% CI 74.3 to 84.4); asked whether the problems were sufficiently demanding to be worthwhile, 66.8\% agreed (95\% CI 60.6 to 72.5). These are perceptions rather than measures of learning. Full distributions for the six Likert items are shown in Fig.~\ref{fig:likert} and Table~\ref{tab:likert}.

\begin{table}[tbp]
\centering
\caption{Responses to the peer validation Likert items (n = 238; Q25 n = 237)}
\label{tab:likert}
\small
\setlength{\tabcolsep}{4pt}
\begin{tabular}{>{\raggedright\arraybackslash}p{5.2cm}rrrrrp{3.0cm}rp{1.0cm}}
\toprule
\textbf{Item} & \textbf{SD} & \textbf{D} & \textbf{N} & \textbf{A} & \textbf{SA} & \textbf{\% agree [95\% CI]} & \textbf{Mdn} & \textbf{IQR} \\
\midrule
Q21. Solved problems I could not solve before & 3 & 3 & 42 & 116 & 74 & 79.8 [74.3, 84.4] & 4 & 4--5 \\
Q22. Problems were hard enough to be worth doing & 4 & 6 & 69 & 111 & 48 & 66.8 [60.6, 72.5] & 4 & 3--4 \\
Q23. Peer validation made me put in less effort [R] & 40 & 88 & 67 & 37 & 6 & 18.1 [13.7, 23.4] & 2 & 2--3 \\
Q24. I validated the same person who validated me [R] & 68 & 76 & 62 & 26 & 6 & 13.4 [9.7, 18.4] & 2 & 1--3 \\
Q25. My validator applied a lower bar [R] & 38 & 62 & 99 & 32 & 6 & 16.0 [11.9, 21.2] & 3 & 2--3 \\
Q26. Trusted that others validated honestly & 5 & 13 & 61 & 112 & 47 & 66.8 [60.6, 72.5] & 4 & 3--4 \\
\bottomrule
\end{tabular}

\vspace{0.35em}
\footnotesize SD = strongly disagree, D = disagree, N = neutral, A = agree, SA = strongly agree. Cell entries are counts. Intervals are Wilson intervals for the proportion agreeing.
\end{table}

\textbf{Reported integrity problems (RQ1).} Three items asked participants about their own conduct and experience. Agreement was 18.1\% for having reduced effort (95\% CI 13.7 to 23.4), 16.0\% for the validator having applied a lower bar than the teaching team (95\% CI 11.9 to 21.2), and 13.4\% for having validated the person who validated them (95\% CI 9.7 to 18.4). Sixty-nine respondents, or 29.0\% (95\% CI 23.6 to 35.1), agreed with at least one of these items; of these, 39 agreed with one, 16 with two, and 14 with all three. The use of the midpoint was substantial (28.2\%, 41.8\%, and 26.1\%, respectively), and neutral responses were therefore not combined with either tail. The reciprocity item, unlike the other two, concerns behaviour that was prohibited by programme rule, and its figure of 13.4\% is best interpreted as a lower bound.

\textbf{Perceptions of others compared with self (RQ2).} Asked what proportion of validations across the programme had been granted without proper examination, only 11.3\% selected ``almost none''; 88.7\% selected at least ``a few'', and 37.8\% (95\% CI 31.9 to 44.1) selected ``about a quarter'' or more, as shown in Table~\ref{tab:laxity}. This stands alongside the 13.4\% who reported reciprocal validation on their own part.

\begin{table}[tbp]
\centering
\caption{Perceived proportion of lax validation and ease of finding a validator (n = 238)}
\label{tab:laxity}
\begin{tabularx}{0.9\linewidth}{Y r r l}
\toprule
\textbf{Response} & \textbf{n} & \textbf{\%} & \textbf{95\% CI} \\
\midrule
\multicolumn{4}{l}{\textbf{Q27. Proportion validated without proper examination}}\\
Almost none & 27 & 11.3 & [7.9, 16.0] \\
A few & 121 & 50.8 & [44.5, 57.1] \\
About a quarter & 47 & 19.7 & [15.2, 25.3] \\
About half & 24 & 10.1 & [6.9, 14.6] \\
Most & 19 & 8.0 & [5.2, 12.1] \\
\addlinespace
\multicolumn{4}{l}{\textbf{Q28. Ease of finding a validator}}\\
Easy, someone always available & 54 & 22.7 & [17.8, 28.4] \\
Took some chasing & 143 & 60.1 & [53.7, 66.1] \\
Hard, waited a long time & 27 & 11.3 & [7.9, 16.0] \\
Struggled to find anyone & 14 & 5.9 & [3.5, 9.6] \\
\bottomrule
\end{tabularx}
\end{table}

\begin{figure}[tbp]
\centering
\includegraphics[width=0.8\linewidth]{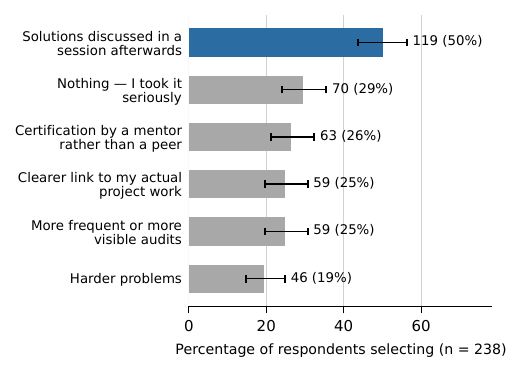}
\caption{Options selected in response to what would make validation more rigorous (multi-select; n = 238). Bars show Wilson 95\% intervals; percentages sum to more than 100 because respondents could select several options.}
\label{fig:rigour}
\end{figure}

\textbf{Changes selected by participants (RQ3).} Asked what would make validation more rigorous, the 238 respondents made 423 selections, as shown in Table~\ref{tab:changes} and Fig.~\ref{fig:rigour}. The discussion of solutions in a session afterwards was the most frequent selection, at 50.0\% (95\% CI 43.7 to 56.3). The next most frequent response was ``nothing, I took it seriously'', at 29.4\%. Validation by a mentor followed at 26.5\%, more visible auditing and a clearer link to project work at 24.8\% each, and more demanding problems at 19.3\%. The options were not selected at equal rates (Cochran's Q(4) = 73.37, p < 0.001). The discussion session was selected roughly twice as often as auditing (ratio 2.02, 95\% CI 1.61 to 2.60) and as mentor validation (ratio 1.89, 95\% CI 1.49 to 2.46), corresponding to paired differences of approximately 25 and 24 percentage points (Table~\ref{tab:pairwise}); auditing and mentor validation did not differ from one another (p = 0.72). Excluding the 70 respondents who selected ``nothing'' leaves the ordering unchanged and widens the margin, with the discussion session rising to 63.1\% of the remaining 168. Demand for the discussion session was uniform across every subgroup examined (46.5\% to 53.3\%, no comparison approaching significance), whereas demand for supervision was concentrated among those who had reported a problem: auditing was selected by 46.9\% of self-reported reciprocal validators, compared with 21.4\% of the remainder (odds ratio 3.25, p = 0.004; these subgroup tests are exploratory).

\begin{table}[tbp]
\centering
\caption{Changes selected as making validation more rigorous (multi-select; n = 238)}
\label{tab:changes}
\begin{tabularx}{0.9\linewidth}{Y r r l}
\toprule
\textbf{Options} & \textbf{n} & \textbf{\%} & \textbf{95\% CI} \\
\midrule
Solutions discussed in a session afterwards & 119 & 50.0 & [43.7, 56.3] \\
Nothing, I took it seriously. & 70 & 29.4 & [24.0, 35.5] \\
Validation by a mentor rather than a peer & 63 & 26.5 & [21.3, 32.4] \\
More frequent or more visible audits & 59 & 24.8 & [19.7, 30.6] \\
Clearer link to my actual project work & 59 & 24.8 & [19.7, 30.6] \\
More demanding problems & 46 & 19.3 & [14.8, 24.8] \\
Other (free text) & 7 & 2.9 & [1.4, 5.9] \\
\bottomrule
\end{tabularx}
\end{table}

\begin{table}[tbp]
\centering
\caption{Pairwise comparisons between options (exact McNemar tests, paired, n = 238)}
\label{tab:pairwise}
\small
\begin{tabularx}{\linewidth}{Y l l l}
\toprule
\textbf{Comparison} & \textbf{Discordant} & \textbf{Difference [95\% CI]} & \textbf{p} \\
\midrule
Solutions vs. audits & 81 / 21 & +25.2 pp [17.5, 32.9] & $<0.001$ \\
Solutions vs. mentor validation & 87 / 31 & +23.5 pp [15.1, 32.0] & $<0.001$ \\
Solutions vs. project link & 79 / 19 & +25.2 pp [17.7, 32.7] & $<0.001$ \\
Solutions vs. ``nothing'' & 106 / 57 & +20.6 pp [10.4, 30.8] & $<0.001$ \\
Mentor validation vs. audits & 37 / 33 & +1.7 pp [$-5.2$, 8.6] & 0.72 \\
\bottomrule
\end{tabularx}

\vspace{0.35em}
\footnotesize Discordant pairs are reported as (first option only) / (second option only). pp = percentage points. Omnibus test across the five substantive options: Cochran's Q(4) = 73.37, p < 0.001.
\end{table}

\textbf{Availability of validators (RQ4).} Finding a validator required effort for most participants: only 22.7\% reported that someone was always available, while 77.3\% (95\% CI 71.6 to 82.2) reported some difficulty and 17.2\% reported the two most difficult levels. This difficulty was not associated with reciprocal validation ($\rho$ = 0.06, p = 0.34) or with reduced effort ($\rho$ = 0.10, p = 0.11), and its association with perceived laxity did not survive correction for multiple testing ($\rho$ = 0.13, Holm p = 0.20). The two concerns that did move together, reduced effort and a lowered bar ($\rho$ = 0.62), appear to reflect a single underlying impression that the process lacked rigour, rather than two independent observations.

\subsection{Discussion}

The central result runs counter to what an integrity-focused framing might predict. Offered two means of supervising validation more closely and one means of teaching the material more fully, participants selected the instructional option roughly twice as often as either supervisory one, and did so largely irrespective of their own experience. Demand for supervision came disproportionately from those who had reported a lapse, whereas demand for the discussion session was uniform across the sample. This inverts the common instinct to meet suspected dishonesty with closer monitoring, and it is consistent with a substantial body of work arguing that assessment is most valuable when it develops the learner's own capacity to judge quality \citep{Sadler1989,Nicol2006} rather than when it functions primarily to police output.

The free-text responses suggest why. The recurring concern was not that peers behaved dishonestly, but that validation could take place without the concepts being genuinely examined. One participant described questions being treated as ``just explanation and moving on'', and another observed that a certain depth of learning was lost when a peer validated without teaching. In the language of feedback literacy \citep{Carless2018}, the process delivered a verdict without establishing the conditions under which a learner forms and applies a judgement about their own work; the reviewer-side benefits that peer assessment is known to yield arise precisely through that act of judging \citep{Topping1998,Lundstrom2009}. What participants requested, in asking for solutions to be discussed, was the instruction that validation alone had omitted.

This distinction has practical significance, because the two forms of remedy differ markedly in cost. Supervision does not scale, and it consumes the very expert time that peer validation was introduced to conserve; routing validations through mentors, in a programme with three qualified assessors, approaches a contradiction. A single session in which solutions are discussed afterwards reaches an entire cohort at once and constitutes instruction as much as assurance. We cannot demonstrate from these data that such a session would reduce lax validation, and we do not claim as much; what we can report is that it is what participants requested, and that they requested it regardless of whether they had reported a lapse themselves.

A second theme concerns trust as a property distinct from accuracy. A large majority believed that some validations were granted without proper examination, while only a small minority reported having done so themselves. The two items pose different questions and neither verifies the other: both are self-reported, but the item concerning one's own conduct carries a cost of admission that the item concerning others does not, and we therefore interpret the gap as a fact about perception rather than a measure of true prevalence. Its consequence holds in either case. An assessment arrangement whose own participants doubt its outputs is not delivering assurance, whatever the actual rate of lax validation. The peer assessment literature has largely validated the practice against agreement with instructor marks \citep{Falchikov2000,Kulkarni2013}; perceived legitimacy is a different property, is not captured by that measure, and, in an arrangement without an instructor mark, may be the only property that can be observed at all. That a non-anonymous, face-to-face scheme should raise such concerns is consistent with evidence that the social configuration of peer assessment shapes both its candour and its perception \citep{Panadero2019}.

Considered against the wider evidence base, the reassuring finding is that a distributed, instructor-light validation system did not collapse into collusion. Reported reciprocity was low, echoing \citeauthor{Magin2001}'s (\citeyear{Magin2001}) finding that reciprocity contributes little to peer-assessment bias, even in a setting where none of the usual safeguards of aggregation, moderation, and calibration were present \citep{Boud1999,Boud2001}. The difficulty was subtler and more instructive: not dishonesty but insufficiency, validation without teaching. For a community increasingly reliant on peer and technology-mediated assessment at scale \citep{Piech2013,Indriasari2020,Li2010,VanPopta2017}, the implication is that scaling the judgement is not the same as scaling the learning, and that participants themselves are able to perceive the difference.

\subsection{Limitations}

The limitation that governs every claim made here is straightforward: the study relies on self-report from a single cohort at one point in time, with no record of which validator assessed which intern and no direct measure of learning. It can therefore report what participants perceived and preferred, but not whether validation was in fact accurate or whether the activity improved understanding. The single most consequential omission is the validation graph: had it been recorded which validator assessed which intern, it would have been possible to test the attenuation that participants most vividly described, whereby rigour diminished with distance from the instructor along the cascade. That, more than any individual caveat, is what we would advise a subsequent implementation to instrument from the outset.

\section{Conclusion}

Where qualified assessors are scarce, peer validation is less a matter of preference than of necessity, and the pertinent question is not whether to employ it but how it behaves once it performs work that an instructor cannot. This paper has described one large implementation: a cascading peer validation system in which the authority to validate propagated through the cohort from a single instructor, without a confirmation step and without a numeric mark. Among the 238 consenting participants surveyed, most regarded the activity as worthwhile, a minority reported that the standard had slipped, and a large majority suspected lapses in others that they did not acknowledge in themselves. Difficulty in finding a validator was widespread but was not associated with any of the reported problems. Asked what would help, participants chose instruction over supervision by a factor of roughly two, and did so whether or not they had reported a lapse. If there is a single message for others designing peer validation at scale, it is that participants experienced the shortcomings of validation as a shortcoming of teaching, and asked, before anything else, to be taught. Whether addressing that gap through a shared discussion of solutions would in fact raise the standard is the question we would most wish a future study, one that records the validation graph and compares cohorts with and without such a session, to pursue.

\section*{Acknowledgments}

The authors thank the VicharanaShala Lab for Education Design (VLED), Indian Institute of Technology Ropar, for the opportunity to carry out this research within its internship program, for access to the anonymised operational data, and for supporting the research reported here. We also thank the program's operations team and the interns who responded to the survey.

\section*{Disclosure of Interests}

The authors have no competing interests to declare that are relevant to the content of this article. The authors used a generative AI assistant to help restructure and edit drafts of the text and to format the manuscript. The authors designed the study, checked all analyses, results, and references, and take full responsibility for the content.

\bibliographystyle{plainnat}
\bibliography{ref}

\end{document}